\documentclass[aps,pra,reprint,superscriptaddress,twocolumn]{revtex4-2}
\usepackage{adjustbox}

\usepackage[english]{babel}
\usepackage{ucs}
\usepackage[utf8x]{inputenc}
\usepackage{graphicx}
\usepackage{dcolumn}
\usepackage{bm}
\usepackage{bbm}
\usepackage{amsmath}
\usepackage{amssymb}
\usepackage{mathrsfs}
\usepackage{bm}
\usepackage{bbold}
\usepackage{empheq}
\usepackage{amsfonts}
\usepackage{color,colordvi}
\usepackage{physics}
\usepackage[colorlinks=true]{hyperref}
\usepackage[capitalize]{cleveref}
\usepackage{lipsum}
\usepackage{etoolbox}
\usepackage{comment}

\usepackage{xcolor}
\usepackage[normalem]{ulem}

\begin{document}
    \title{Quantum Fisher information analysis for absorption measurements with undetected photons}

	\author{Martin~Houde}
	\affiliation{Department of Engineering Physics, École polytechnique de Montréal, Montréal, QC, H3T 1J4, Canada}
    
    \author{Franz~Roeder}
	\affiliation{Paderborn University, Integrated Quantum Optics and Institute for Photonic Quantum Systems (PhoQS), 33095, Paderborn, Germany}

    \author{Christine~Silberhorn}
	\affiliation{Paderborn University, Integrated Quantum Optics and Institute for Photonic Quantum Systems (PhoQS), 33095, Paderborn, Germany}
    
    \author{Benjamin~Brecht}
	\affiliation{Paderborn University, Integrated Quantum Optics and Institute for Photonic Quantum Systems (PhoQS), 33095, Paderborn, Germany}

    \author{Nicolás~Quesada}
	\affiliation{Department of Engineering Physics, École polytechnique de Montréal, Montréal, QC, H3T 1J4, Canada}
	
	\begin{abstract}
        We theoretically compare the quantum Fisher information (QFI) for three configurations of absorption spectroscopy with undetected idler photons: an SU(1,1) interferometer with inter-source idler loss, an induced-coherence (IC) setup in which the idler partially seeds a second squeezer together with a vacuum ancilla, and a distributed-loss (DL) scheme with in-medium attenuation. We calculate the QFI as a function of parametric gain for both full and signal-only detection access. For losses below 99\% and low to moderate gain, the SU(1,1) configuration provides the largest QFI. At high gain and intermediate loss, the IC scheme performs best, while under extreme attenuation (transmission $<$ 1\%) the DL model becomes optimal. These results delineate the measurement regimes in which each architecture is optimal in terms of information theory.
	\end{abstract}
    \maketitle
	\section{Introduction}
	\label{sec:Intro}

    Measurements with undetected photons \cite{Lindner2021,Kaufmann2022,Neves2024,Paterova2018,Dong2025} have emerged as a powerful approach for quantum spectroscopy at otherwise inaccessible wavelengths \cite{Lindner2021,Haase2023,Kutas2025}, such as in the mid-infrared (MIR) spectral range. The most widely used implementation of these measurements is based on nonlinear SU(1,1) interferometers \cite{Yurke1986}. In such systems, a strongly frequency–time entangled photon-pair source—typically parametric down-conversion (PDC)—generates highly nondegenerate signal and idler photons. By convention, the idler photon probes the object of interest, while the signal, which lies in the visible or near-infrared spectral range, can be efficiently detected. After the idler interacts with the sample, both photons are recombined at a second nonlinear source, where nonlinear interference transfers the information carried by the idler onto the signal, enabling its optical readout.
    
    Recent work has demonstrated that when combined with photon counting, SU(1,1) interferometers provide an unconditional quantum metrological advantage over conventional Mach–Zehnder interferometers \cite{Santandrea2023}. In the single photon-pair generation regime, it has been shown that SU(1,1) interferometers combined with intensity difference measurements can provide a loss-tolerant advantage for absorption sensitivity compared to classical schemes \cite{Okamoto2020Loss}. From an experimental point of view, SU(1,1) interferometers have become the central platform for broadband quantum spectroscopy \cite{Lindner2021,Panahiyan2022,Panahiyan2023,Neves2024,Kaufmann2022,Riazi2019dispersion}.
    
    Two additional experimental configurations have been developed to realize measurements with undetected photons. The induced-coherence (IC) scheme \cite{Wang91} closely resembles the SU(1,1) configuration but differs in that only the idler from the first source propagates through the second. A new signal photon is generated there that carries the encoded information from the idler. Interference between the two signal modes on a beam splitter then reveals the sample-dependent changes. The IC configuration enabled the first experimental realization of measurements with undetected photons \cite{Lemos2014} and has recently attracted renewed attention \cite{Gemmell2024,Roeder_OCT, volkoff2024radar}.
    
    In contrast, a more recent approach exploits the interaction of the idler with the analyte during the generation process itself \cite{Kumar2020,Krstic2023}. In this distributed-loss (DL) configuration, the idler experiences loss while the photon pair is generated, which imprints the corresponding absorption information onto the signal spectrum. This method eliminates the need for nonlinear interference and thereby simplifies the experimental setup, but requires analytes that can interact with the nonlinear medium—such as liquids or gases within or near a waveguide—and is thus less universally applicable.
    
    While numerous experimental applications of measurements with undetected photons have been reported, their fundamental information-theoretic capabilities remain underexplored \cite{Boyd2017}, particularly in terms of the quantum Fisher information (QFI). The QFI quantifies the ultimate information attainable about a parameter from a given quantum state and determines the theoretical bound on achievable estimation precision \cite{Helstrom1969,Holevo1982ProbabilisticAS,Braunstein1994QFIseminal,Paris2009QFIseminal}. Although the QFI in SU(1,1)-type interferometers has been analyzed in the low-gain limit, the influence of high parametric gain on the relative performance of the SU(1,1), IC, and DL configurations remains unclear. Previous studies indicate that the sensitivity of nonlinear interferometers can increase at high gain \cite{Sharapova2018,Scharwald2023,Hashimoto2024}, and that gain dependence also manifests in distributed-loss absorption measurements \cite{Krstic2023}. However, a comprehensive quantitative comparison is still lacking.
    
    In this work, we develop a theoretical framework for broadband photon-pair sources based on spatial Heisenberg–Langevin equations of motion. This approach enables the calculation of frequency-resolved second-order photon-number moments for arbitrary dispersion and gain, from which we derive the QFI for absorption spectroscopy with undetected photons across the SU(1,1), IC, and DL schemes. We find that, across most experimentally relevant regimes—encompassing low to moderate gain and losses below 99\%—the SU(1,1) configuration yields the highest QFI. At high gain, however, the IC setup exhibits superior performance over a limited loss range, while the DL configuration becomes optimal for extreme loss exceeding 99\%.
    
\begin{figure}
		\includegraphics[width=1\linewidth]{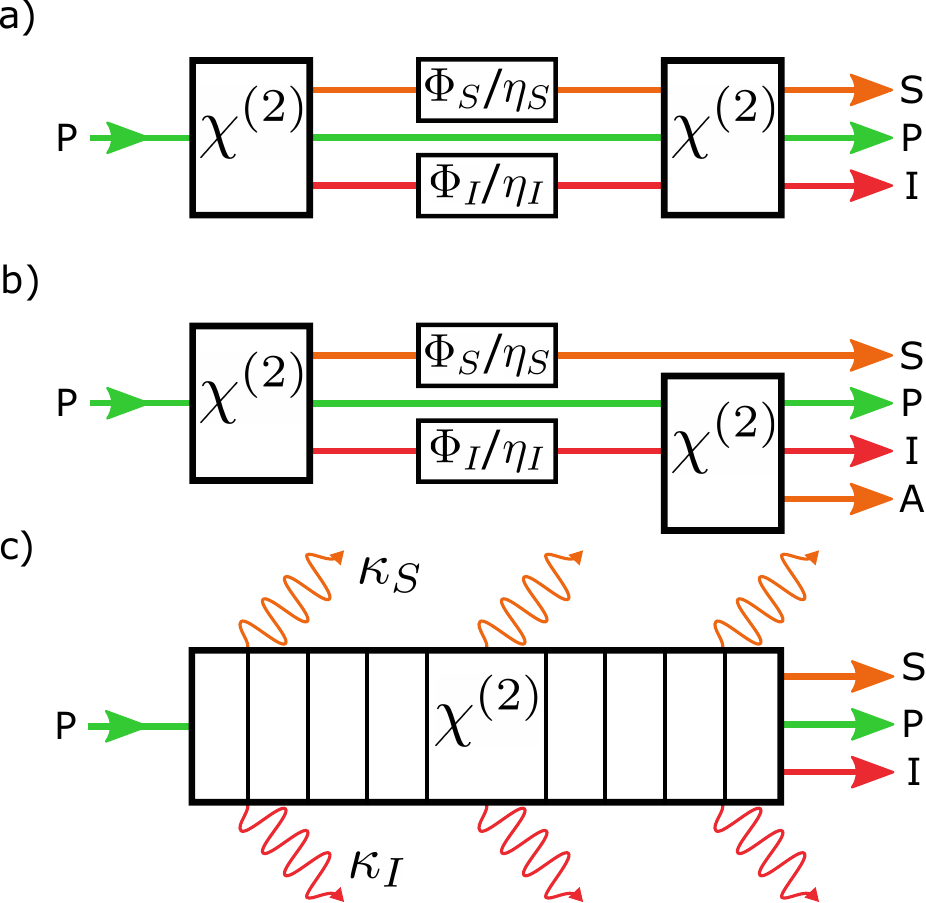}
		\caption{Schematic representation of the different sensing configurations and the respective loss implementations. (a) SU(1,1) interferometer where the two nonlinear regions are taken to be identical and lossless. We treat losses via a beam splitter interaction (with transmission rates $\eta_{S}$ and $\eta_{I}$ for the signal and idler respectively) between both nonlinear regions. (b) IC system where in the second nonlinear region the idler is mixed with vacuum of an ancilla mode. (c) DL model where the signal and idler modes experience different decay rates ($\kappa_{S}$ and $\kappa_{I}$ respectively) as they propagate through the nonlinear region. To simulate optical path delays, we also allow the beam splitter interaction to induce additional dispersion for the signal and idler modes ($\Phi_{S}$ and $\Phi_{I}$ respectively) in the SU(1,1) and IC configuration. Both losses and transmission rates are taken to be frequency dependent. P: pump; S: signal; I: idler; A: ancilla.}
		\label{fig:model}
	\end{figure}

    The remainder of this paper is organized as follows. In Sec.~\ref{sec:cw}, we present the theoretical model for twin-beam generation in the continuous-wave limit and obtain frequency-resolved second-order moments. Analytical results for the SU(1,1), IC, and DL setups—depicted in Figs.~\ref{fig:model}(a–c)—are derived in Secs.~\ref{sec:su11}, \ref{sec:IC}, and \ref{sec:dist}, respectively. In Sec.~\ref{sec:param}, we briefly discuss how we parametrize the losses and the levels of gain in the different models in a consistent manner. We introduce the concept of the QFI and its role as a figure of merit for absorption estimation in Sec.~\ref{sec:QFI}, followed by a full multimode QFI analysis in Sec.~\ref{sec:fullQFI}. In Sec.~\ref{sec:signalQFI}, we consider the QFI when we do not have access to the idler, specifying the case of the IC configuration with access to both signal and ancilla in Sec.~\ref{sec:IC_twomode}  and then the case of single-mode operation in Sec.~\ref{sec:single}. In Sec.~\ref{sec:analysis} we compare the different QFI results when we do not have idler access. Finally, we summarize our main findings in Sec.~\ref{sec:conc}.

	\section{Theory model}
	\label{sec:cw}
    \subsection{Lossless continuous wave model}
    We begin by considering a lossless continuous-wave model for the nonlinear interaction. We use the ($z$, $\omega$) equations of motion for twin-beam generation derived and used in~\cite{christ2013theory,houde2023sources,quesada2020theory,lipfert2018bloch,kolobov1999spatial,horoshoko2022generator}. For each mode, $j=P,S,I$ for pump, signal, and idler respectively, we associate a central wavevector $\bar{k}_{j}$ with a central frequency $\bar{\omega}_{j}$ such that $\bar{k}_{j} = k_{j}(\bar{\omega}_{j})$ where $k_{j}(\omega)$ is the dispersion relation determined by the material and geometry of the squeezer. For type-II SPDC to occur, we require that
    \begin{align}
		\bar{\omega}_{P}-\bar{\omega}_{S}-\bar{\omega}_{I}&=0,\label{Eq:EnconMT}\\
		\bar{k}_{P}-\bar{k}_{S}-\bar{k}_{I}&=0.\label{Eq:MomconMT}
	\end{align}
	Note that if quasi-phase matching is used, the right-hand side of Eq.~(\ref{Eq:MomconMT}) should be changed to $\pm 2\pi/\Lambda_{\text{pol}}$ where $\Lambda_{\text{pol}}$ is the poling period. For these equations to be valid, we assume that self- and cross-phase modulation terms are negligible and assume that the pump mode is prepared in a strong coherent state with a large number of photons which remains constant throughout the interaction (undepleted-classical pump approximation)\cite{Chinni2024pump}. Furthermore, we take the continuous-wave monochromatic pump limit where the pump spectral profile becomes proportional to a delta function centered at the pump's central frequency.  Under these assumptions, the resulting equations of motion for the signal and idler modes are
    \begin{align}
     \frac{\partial}{\partial z}\hat{c}_{S}(\omega,z)=i\Delta k_{S}(\omega) \hat{c}_{S}(\omega,z) + i\gamma \hat{c}^{\dagger}_{I}(-\omega,z),\label{eq:eom_cws}\\
     \frac{\partial}{\partial z}\hat{c}^{\dagger}_{I}(\omega,z)=-i\Delta k_{I}(\omega) \hat{c}^{\dagger}_{I}(\omega,z) - i\gamma^* \hat{c}_{S}(-\omega,z)\label{eq:eom_cwi}.
    \end{align}
    where $\Delta k_{S/I}(\omega) = k_{S/I}(\omega) - \bar{k}_{S/I}$ are the dispersion relations detuned from their central wavevectors, $\gamma=|\gamma|e^{-i\Phi_{P}}$ represents the interaction strength which depends on the phase of the pump and many other material parameters~\cite{quesada2020theory,quesada2022BPP,horoshoko2022generator,helt2020degenerate}, $x^{*}$ denotes the complex conjugate of $x$. In all the equations in this manuscript the variable $\omega$ is used to indicate a detuning from a central frequency. Thus, for example, $\hat{a}_S(\omega,z)$ denotes a Heisenberg evolved operator that destroys  photons of frequency $\bar{\omega}_S + \omega$; similarly, $k_I(-\omega)$, describes the wavevector associated with frequency $\bar{\omega}_I - \omega$. This notation is necessary to avoid unnecessary cluttered equations. The equations of motion thus link the signal annihilation operator at frequency $\bar{\omega}_{S}+\omega$ to the idler creation operator at $\bar{\omega}_{I}-\omega$, a consequence of energy conservation, see Eq.~(\ref{Eq:EnconMT}). 

    For numerical evaluation, we discretize frequency space such that $\omega_{n} = \omega_{0}+n\Delta\omega|^{N-1}_{0}$ for an $N$-size grid and introduce new, dimensionless operators $\hat{a}_{S/I}(\omega_{n}) = \hat{c}_{S/I}(\omega_{n})\sqrt{\Delta \omega}$. These new operators obey the same Heisenberg equations of motion (Eqs.(\ref{eq:eom_cws}) and (\ref{eq:eom_cwi}))
    
    Since the equations of motions are $z$-independent, we can solve them analytically by matrix exponentiation. To do so, we first express the equations of motion in matrix form
    \begin{align}
    \frac{\partial}{\partial z}\begin{pmatrix}
        \hat{a}_{S}(\omega,z) \\
        \hat{a}^{\dagger}_{I}(-\omega,z)
    \end{pmatrix} &= i\bm{Q}(\omega)\begin{pmatrix}
        \hat{a}_{S}(\omega,z) \\
        \hat{a}^{\dagger}_{I}(-\omega,z)
    \end{pmatrix}\nonumber\\ &= i\begin{pmatrix}
        \Delta k_{S}(\omega) & \gamma \\
        -\gamma^* & -\Delta k_{I}(-\omega)
    \end{pmatrix}\begin{pmatrix}
        \hat{a}_{S}(\omega,z) \\
        \hat{a}^{\dagger}_{I}(-\omega,z)
    \end{pmatrix}.
    \end{align}
    The solution is given by the propagator $\bm{U}(\omega;z-z_{0})$, also known as a transfer matrix (in the continuous-wave case), such that
    \begin{align}\label{eq:matrix_eom}
        \begin{pmatrix}
        \hat{a}_{S}(\omega,z) \\
        \hat{a}^{\dagger}_{I}(-\omega,z)
        \end{pmatrix} = \bm{U}(\omega;z-z_{0})\begin{pmatrix}
        \hat{a}_{S}(\omega,z_{0}) \\
        \hat{a}^{\dagger}_{I}(-\omega,z_{0})\end{pmatrix}
    \end{align}
    where
    \begin{align}\label{eq:propagator}
        \bm{U}(\omega;z-z_{0})&=e^{i\bm{Q}(\omega)(z-z_{0})}\nonumber\\
        &\equiv\begin{pmatrix}
           \bm{U}^{S,S}(\omega;z-z_{0}) & \bm{U}^{S,I}(\omega;z-z_{0})  \\
           \bm{U}^{I,S}(\omega;z-z_{0}) & \bm{U}^{I,I}(\omega;z-z_{0})
        \end{pmatrix}
    \end{align}
    with matrix elements
    \begin{widetext}
    \begin{eqnarray}
    \bm{U}^{S,S}(\omega;z-z_{0}) &=& e^{i\Delta K(\omega) (z-z_{0})/2}\left[ \cos{\left(\nu(\omega) (z-z_{0})/2\right)}\vphantom{\frac{\Sigma_{K}(\omega)}{\nu(\omega)}}+i\frac{\Sigma_{K}(\omega)}{\nu(\omega)}\sin{(\nu(\omega) (z-z_{0})/2)}    \right],\label{eq:uss}   \\
    \bm{U}^{I,I}(\omega;z-z_{0}) &=& e^{i\Delta K(\omega) (z-z_{0})/2}\left[ \cos{\left(\nu(\omega) (z-z_{0})/2\right)}\vphantom{\frac{\Sigma_{K}(\omega)}{\nu(\omega)}}-i\frac{\Sigma_{K}(\omega)}{\nu(\omega)}\sin{(\nu(\omega) (z-z_{0})/2)}    \right],\label{eq:uii}\\
    \bm{U}^{S,I}(\omega;z-z_{0}) &=&-\frac{\gamma}{\gamma^*}\bm{U}^{I,S}(\omega;z-z_{0})= 2i\gamma e^{i\Delta K(\omega) (z-z_{0}))/2} \frac{\sin{(\nu(\omega) (z-z_{0})/2)}}{\nu(\omega)} , \label{eq:usi}
    \end{eqnarray}
    \end{widetext}
    where we have defined
    \begin{align}
        \Delta K(\omega) &= \Delta k_{S}(\omega) -\Delta k_{I}(-\omega),\\
        \Sigma_{K}(\omega) &=\Delta k_{S}(\omega) +\Delta k_{I}(-\omega),\\
        \nu(\omega) &= \sqrt{\left[\Sigma_{K}(\omega)\right]^2 -4|\gamma|^{2}}\label{eq:nuK}.
    \end{align}
    With the solutions in hand, we can evaluate any second order moment after the nonlinear interaction as a function of these propagator matrix elements and the second order moments before the interaction. 
    Before we obtain the general expression for second order moments, we begin by obtaining expressions when the modes before the interaction, the input modes, are in vacuum. Without loss of generality, we set $z_{0}=0$ and take the interaction region to be of length $L=40$ mm. In terms of the propagator matrix elements, we find that the second order moments for the signal and idler modes after passing through the nonlinear region, with respect to (w.r.t.) vacuum input modes, are
    \begin{widetext}
    \begin{subequations}\label{eq:vac_mom}
    \begin{eqnarray}
        N^{\text{V}}_{S}(\omega) &=& \langle\hat{a}^{\dagger}_{S}(\omega,L)\hat{a}_{S}(\omega,L) \rangle_{ \text{vac}} = \bm{U}^{S,I}(\omega;L)\left[\bm{U}^{S,I}(\omega;L)\right]^{*},\label{eq:Nsv} \\
        N^{\text{V}}_{I}(-\omega) &=& \langle  \hat{a}^{\dagger}_{I}(-\omega,L)\hat{a}_{I}(-\omega,L) \rangle_\text{vac} = \bm{U}^{I,S}(\omega;L) \left[\bm{U}^{I,S}(\omega;L)\right]^{*},\label{eq:Niv}\\
        M^{\text{V}}(\omega) &=& \langle \hat{a}_{S}(\omega,L)\hat{a}_{I}(-\omega,L) \rangle_\text{vac} = \bm{U}^{S,S}(\omega;L)\left[\bm{U}^{I,S}(\omega;L)   \right]^{*}\label{eq:Mv}.
    \end{eqnarray}
    \end{subequations}
    \end{widetext}
     where we define the Heisenberg picture expectation value w.r.t. vacuum as $\langle \bm{\cdot}\rangle_{\text{vac}}=\langle\text{vac}|\bm{\cdot}|\text{vac}\rangle$. The superscript ``V'' labels the fact that the expectation values are w.r.t. vacuum input modes. These should not be confused with the second order moments for vacuum modes which are $\langle\hat{a}^{\text{vac}}(\omega,0)\left[\hat{a}^{\text{vac}}(\omega,0)\right]^{\dagger} \rangle_{ \text{vac}}=1$ or 0 otherwise. Note that by symmetry, we have that $N^{\text{V}}_{S}(\omega)=N^{\text{V}}_{I}(-\omega)\equiv N^{\text{V}}(\omega)$. Explicitly, we have that
     \begin{align}
         N^{\text{V}}(\omega) =& \left|\frac{2\gamma\sin{(\nu(\omega) L/2)}}{\nu(\omega)}\right|^{2}\label{eq:Nv},\\
         M^{\text{V}}(\omega) =&2i\gamma\left[ \frac{\sin{(\nu(\omega) L/2)}}{\nu(\omega)}  \right]^{*}\biggl[ \cos{(\nu(\omega) L/2)} \nonumber\\
         &\left.+i\frac{\Sigma_{K}(\omega)}{\nu(\omega)}\sin{(\nu(\omega) L/2)}    \right]\label{eq:Mv_exp}.
     \end{align}

    With these definitions, we can then consider a more general case where before the nonlinear interaction the second order moments are non-zero (e.g. second pass in SU(1,1) setup of Fig.~\ref{fig:model}(a)). We label moments before the interaction with superscript ``in'' and those after with superscript ``out''. In the general case, we find that
    \begin{widetext}
    \begin{eqnarray}
        N^{\text{out}}_{S}(\omega) &=& N^{\text{in}}_{S}(\omega)\left( 1+N^{\text{V}}(\omega) \right) +N^{\text{V}}(\omega)\left( 1+N^{\text{in}}_{I}(-\omega) \right)  -2\text{Re}\left[\frac{\gamma^{*}}{\gamma} M^{\text{V}}(\omega)M^{\text{in}}(\omega)  \right],\label{eq:ns_out}\\
        N^{\text{out}}_{I}(-\omega) &=& N^{\text{in}}_{I}(-\omega)\left( 1+N^{\text{V}}(\omega) \right) +N^{\text{V}}(\omega)\left( 1+N^{\text{in}}_{S}(\omega) \right) -2\text{Re}\left[ \frac{\gamma^{*}}{\gamma}M^{\text{V}}(\omega)M^{\text{in}}(\omega)  \right],\label{eq:ni_out}\\
        M^{\text{out}}(\omega) &=& M^{\text{V}}(\omega)\left( 1+N^{\text{in}}_{S}(\omega) \right) +M^{\text{V}}(\omega)N^{\text{in}}_{I}(-\omega) -\frac{\gamma}{\gamma^*}N^{\text{V}}(\omega)\left[M^{\text{in}}(\omega)\right]^* -\frac{\gamma^{*}}{\gamma}\frac{\left[   M^{\text{V}}(\omega)\right]^2}{N^{\text{V}}(\omega)}M^{\text{in}}(\omega).\label{eq:m_out}
    \end{eqnarray}
    \end{widetext}
    See Appendix~\ref{app:moments} for an in depth derivation of all the above moments. With these expressions, we can now consider the SU(1,1) model with lossless sources. Note that throughout the manuscript we also refer to the second order moment $N_{S}(\omega)$($N_{I}(\omega)$) as the signal(idler) intensity.

	\section{SU(1,1) Model}
    \label{sec:su11}
	For the SU(1,1) model, we treat the sources as ideal squeezers without loss and allow for the modes to incur loss and additional dispersion in between these ideal sources. Furthermore, we consider the case where the two squeezers are perfectly mode matched \cite{houde2024perfect}, i.e. where the input and output modes of the squeezers are identical. Misalignment of the squeezers could lead to distinguishability between the modes of the two squeezers which requires a completely different analysis \cite{volkoff2025su11} which is outside the scope of this manuscript. We derive expressions for the moments in full generality and specify the case of idler only loss after. From the previous section, we know how the second order moments are modified when passing through the nonlinear region for both vacuum and general inputs. We now need to find expressions for how the moments are modified by additional dispersion and loss. To do so, we treat the loss and additional dispersion via beamsplitter interactions (with the bath modes taken to be Markovian quantum noise). In terms of the moments before the additional dispersion and loss (in) we find that the output moments (out) are given by
    \begin{align}
        N^{\text{out,BS}}_{S}(\omega)&=\eta_{S}(\omega)N^{\text{in}}_{S}(\omega)\\
        N^{\text{out,BS}}_{I}(-\omega)&=\eta_{I}(-\omega)N^{\text{in}}_{I}(-\omega)\\
        M^{\text{out,BS}}(\omega)&=\sqrt{\eta_{S}(\omega)\eta_{I}(-\omega)}e^{i\left( \Phi_{S}(\omega)+\Phi_{I}(-\omega)    \right)}M^{\text{in}}(\omega)
    \end{align}
    where $\eta_{S}(\omega)/\eta_{I}(-\omega)$ are frequency dependent transmission coefficients (which we can relate to decay rates), $\Phi_{S}(\omega)/\Phi_{I}(-\omega)$ are additional dispersions (from possible optical path delays or external dispersive elements), and the superscript ``BS'' is to specify that we model a beamsplitter interaction.
    
    We can now construct the solution to the moments after the full SU(1,1) evolution presented in Fig.~\ref{fig:model}(a) in a modular way. For the first pass, we take the input modes to be vacuum and set $\Phi_{P}=0$  to set a reference phase, these are then fed through beamsplitter interactions, which are then in turn fed into a second pass. Note that for the second pass we let $\Phi_{P}$ be arbitrary while keeping everything else identical. This allows one to tune the phases accordingly to operate the second pass as an anti-squeezer. Combining everything together, we find that the second order moments after the SU(1,1) interaction are given by
    \begin{widetext}
    \begin{align}
        N^{\text{SU(1,1)}}_{S}(\omega)=&N^{\text{V}}(\omega)(1+\eta_{S}(\omega))+\left[N^{\text{V}}(\omega)\right]^{2}(\eta_{S}(\omega)+\eta_{I}(-\omega)) \nonumber\\&-2\sqrt{\eta_{S}(\omega)\eta_{I}(-\omega)}\text{Re}\left[e^{i(\Phi_{S}(\omega)+\Phi_{I}(-\omega)-\Phi_{P})}\left[ M^{\text{V}}(\omega) \right]^{2}\right],\label{eq:ns_su11}\\
        N^{\text{SU(1,1)}}_{I}(-\omega)=&N^{\text{V}}(\omega)(1+\eta_{I}(-\omega))+\left[N^{\text{V}}(\omega)\right]^{2}(\eta_{S}(\omega)+\eta_{I}(-\omega)) \nonumber\\&-2\sqrt{\eta_{S}(\omega)\eta_{I}(-\omega)}\text{Re}\left[e^{i(\Phi_{S}(\omega)+\Phi_{I}(-\omega)-\Phi_{P})}\left[ M^{\text{V}}(\omega) \right]^{2}   \right], \\
        M^{\text{SU(1,1)}}(\omega)=&e^{-i\Phi_{P}}M^{\text{V}}(\omega)+e^{-i\Phi_{P}}N^{\text{V}}(\omega)M^{\text{V}}(\omega)(\eta_{S}(\omega)+\eta_{I}(-\omega)) \nonumber\\&-\sqrt{\eta_{S}(\omega)\eta_{I}(-\omega)}e^{-i(\Phi_{S}(\omega)+\Phi_{I}(-\omega)+2\Phi_{P})}N^{\text{V}}(\omega)\left[ M^{\text{V}}(\omega) \right]^{*}\nonumber\\&-\sqrt{\eta_{S}(\omega)\eta_{I}(-\omega)}e^{i(\Phi_{S}(\omega)+\Phi_{I}(-\omega))}\frac{\left[ M^{\text{V}}(\omega) \right]^{3}}{N^{\text{V}}(\omega)},   
    \end{align}
    \end{widetext}
    where we have explicitly written out the pump phase dependencies (i.e. it is understood that $\gamma=|\gamma|$ in the second order moments). These solutions will allow us to study the signal spectrum at the output of the SU(1,1) interferometer for different analytes and operation conditions.

    As mentioned previously, when considering parameter estimation with an SU(1,1) configuration, one typically wants to operate the second squeezer as an anti-squeezer\cite{Sahota2015phaseestimation}. To do so we must also take into consideration the additional phases imprinted by the beamsplitter interactions and the phases accrued in the nonlinear media which come from the phase-sensitive moment $M$. Indeed, for anti-squeezing operation we require that $\cos\left[ \Phi_{S}(\omega)+\Phi_{I}(-\omega)-\Phi_{P}+\Psi(\omega)     \right]=-1$, where $\Psi(\omega)$ is the phase contribution from $\left[M^{V}(\omega)\right]^2$. From Eq.~(\ref{eq:Mv_exp}), we find that $\Psi(\omega)\propto \Sigma_{K}(\omega)L\approx 0$ when we are phase-matched. Therefore, we can set the second squeezer to act as anti-squeezer independently of frequency by not adding additional phases and tuning the phase of the pump accordingly(i.e. $\Phi_{P}=\pi$).

	\section{Induced Coherence model}
	\label{sec:IC}
    Similarly to the SU(1,1) model, for the IC model (see Fig.~\ref{fig:model}(b)) we consider two ideal lossless squeezers and allow for the signal and idler modes to incur loss and additional dispersion in between said squeezers. We also follow the same phase conventions for the interaction strength parameter of the two passes. However, unlike the SU(1,1) model, the IC model makes use of a third mode, which we call the ``ancilla" mode labelled with subscript ``A''. For the second pass, we seed the squeezer with the idler mode and allow it to interact with the ancilla in vacuum. Using the nonlinear and beamsplitter update rules mentioned above, we can construct solutions to the second order moments after the second pass of the IC setup. Since we are inducing coherence between the signal and the ancilla, we need to keep track of more second order moments. We find that the intensities are
    \begin{align}
        N^{\text{IC}}_{S}(\omega) &= \eta_{S}(\omega)N^{\text{V}}(\omega),\\
        N^{\text{IC}}_{I}(-\omega) & =N^{\text{V}}(\omega)+\eta_{I}(-\omega)N^{\text{V}}(\omega)(1+N^{\text{V}}(\omega)),\\
        N^{\text{IC}}_{A} (\omega)&= N^{\text{V}}(\omega)(1+\eta_{I}(-\omega)N^{\text{V}}(\omega)).
    \end{align}
    Due to the induced coherence, we find a non-zero mixed second order moment of the form
    \begin{align}\label{eq:Nas}
        N^{\text{IC}}_{S,A}(\omega) =& \langle \hat{a}^{\dagger}_{S}(\omega;L)\hat{a}_{A}(\omega;L)  \rangle_{\text{vac}}\nonumber\\
         =& \sqrt{\eta_{S}(\omega)\eta_{I}(-\omega)}e^{-i\left(\Phi_{S}(\omega)+\Phi_{I}(-\omega)-\Phi_P\right)}\nonumber\\ & \cdot N^{\text{V}}(\omega)U^{I,I}(\omega; L),
    \end{align}
    where $U^{I,I}(\omega; L)$ is given in Eq.~(\ref{eq:uii}). We also find two non-zero phase-sensitive moments, which we now label with two indices to indicate the relevant modes
    \begin{align}
        M^{\text{IC}}_{S,I}(\omega) =& \sqrt{\eta_{S}(\omega)\eta_{I}(-\omega)}e^{i\left(\Phi_{S}(\omega)+\Phi_{I}(-\omega)\right)}\nonumber\\ &\cdot M^{\text{V}}(\omega)\left[U^{I,I}(\omega; L)\right]^{*}\\
        M^{\text{IC}}_{A,I}(\omega) =& e^{-i\Phi_{P}} M^{\text{V}}(\omega)(1+\eta_{I}(-\omega)N^{V}(\omega)).
    \end{align}
    Note that in the expressions above we have again explicitly written out the pump phase dependencies.

    Typically, when considering doing measurements in the IC setup, one recombines the signal and ancilla modes via a balanced 50:50 beamsplitter and then uses either one or both arms to conduct said measurements. For informational purposes it is important to consider the case where we have access to only one of the beamsplitter arms. Labelling the modes at the output of the balanced beamsplitter with subscripts ``$\pm$'', we find that the relevant second order moments are
    \begin{align}\label{eq:bbs_intensity}
        N^{\text{IC}, \text{BBS}}_{\pm}(\omega) = \frac{1}{2}\left[ N^{\text{IC}}_{S}(\omega)+N^{\text{IC}}_{A}(\omega) \pm 2\text{Im}\left[ N^{\text{IC}}_{S,A} (\omega) \right]   \right],
    \end{align}
    where the superscript ``BBS'' is to specify that we have made use of a balanced beamsplitter. Although there are other non-zero moments, when we consider the case where one of the arms is traced out, only this moment is relevant.

    Focusing on a frequency where we are phase-matched ($\Sigma_{K}\approx 0$) and considering idler only loss, we find that the intensities after the balanced beamsplitter (Eq.~(\ref{eq:bbs_intensity})) take the form
    \begin{align}\label{eq:IC_bbs}
        N^{\text{IC,BBS}}_{\pm}=&\frac{1}{2}\biggl[2N^{V}+\eta_{I}\left[N^{V}\right]^{2}    \nonumber\\ &\pm 2\sqrt{\eta_{I}}N^{V}\sqrt{1+N^{V}}\sin(\frac{\Delta K L}{2}-\Phi)\biggr]
    \end{align}
    where we have used the fact that $U^{I,I}(\omega;L) = e^{i\Delta K L/2}\cosh(\gamma L) = e^{i\Delta K L/2}\sqrt{1+N^{V}}$
    when phase-matched. We have also grouped the additional phases $\Phi=\Phi_{S}(\omega)+\Phi_{I}(-\omega)-\Phi_{P}$. As in the SU(1,1) model, one ideally wants to operate the second squeezer as an anti-squeezer. However, unlike the SU(1,1) model, additional phases come in the form of $\Delta KL/2$ which, unlike $\Sigma_{K}L$, can be non-zero for broadband PDC sources \cite{Roeder2024ultrabroad} (i.e. when the source is still phase-matched away from central frequencies). Therefore, for optimal operation one needs to change the phases as a function of frequency to cancel out the additional phases coming from the $\Delta K(\omega)L/2$ term. This adds an experimental level of complication which should be taken into account when comparing the SU(1,1) and IC setups.

	\section{Distributed loss model}
	\label{sec:dist}
	For the DL model (see Fig.~\ref{fig:model}(c)), we follow the derivations of Ref.~\cite{Caves1987quantum} and the similar models used in~\cite{houde2019loss,kopylov2025loss}. For the same assumptions as the models before and in the continuous-wave limit, the spatial quantum Heisenberg-Langevin equations of motion for the signal and idler modes are
    \begin{align}
        \frac{\partial}{\partial z}\hat{a}_{S}(\omega,z)=&i(\Delta k_{S}(\omega) +i\kappa_{S}(\omega)/2)\hat{a}_{S}(\omega,z) \nonumber\\&+ i\gamma \hat{a}^{\dagger}_{I}(-\omega,z)+\sqrt{\kappa_{S}(\omega)}\hat{b}_{S}(\omega,z),\label{eq:eom_s_d}\\
        \frac{\partial}{\partial z}\hat{a}^{\dagger}_{I}(-\omega,z)=&-i(\Delta k_{I}(-\omega)-i\kappa_{I}(-\omega)/2 )\hat{a}^{\dagger}_{I}(-\omega,z) \nonumber\\& - i\gamma^{*} \hat{a}_{S}(\omega,z)+\sqrt{\kappa_{I}(-\omega)}\hat{b}^{\dagger}_{I}(-\omega,z)\label{eq:eom_i_d},
    \end{align}
    where we now have decay rates $\kappa_{S/I}(\omega)$ and associated bath modes $b_{S/I}(\omega,z)$ which we take to be Markovian quantum noise(i.e. $\langle b_{S/I}(\omega) \rangle=0$ and $\langle b_{S/I}(\omega,z)b^{\dagger}_{S/I}(\omega',z') \rangle=\delta(\omega-\omega')\delta(z-z')$). Note that these decay rates have units of inverse length. To solve these equations of motion, we employ a Green's-function-method formalism as in Ref.~\cite{Caves1987quantum}. Without the bath modes, the equation of motions can be solved via matrix exponentiation 
    \begin{align}
        &\begin{pmatrix}
            \hat{a}_{S}(\omega,z) \\
            \hat{a}^{\dagger}_{I}(-\omega,z)
        \end{pmatrix} = \bm{U}^{\text{Dist}}(\omega;z-z_{0})\begin{pmatrix}
            \hat{a}_{S}(\omega,z_0) \\
            \hat{a}^{\dagger}_{I}(-\omega,z_0)
        \end{pmatrix}\nonumber\\
        &=\exp\left[i\begin{pmatrix} 
            K_{S}(\omega)  & \gamma \\
            -\gamma^{*}  &  -K_{I}(\omega)
        \end{pmatrix}\right]\begin{pmatrix}
             \hat{a}_{S}(\omega,z_0) \\
            \hat{a}^{\dagger}_{I}(-\omega,z_0)
        \end{pmatrix},
    \end{align}
    where now $K_{S}(\omega) = \Delta k_{S}(\omega) +i\kappa_{S}/2$ and $K_{I}(-\omega)=\Delta k_{I}(-\omega) -i\kappa_{I}(-\omega)/2$. Note that this new propagator is very similar to the lossless propagator of Eq.~(\ref{eq:propagator}). Indeed, we can obtain the DL propagator matrix elements by replacing $\Delta k_{S}(\omega)\rightarrow K_{S}(\omega)$ and $\Delta k_{I}(-\omega)\rightarrow K_{I}(-\omega)$ in Eqs.~(\ref{eq:uss})~--~(\ref{eq:nuK}). We can construct the full solution to the equations of motion as
    \begin{align}
        &\begin{pmatrix}
            \hat{a}_{S}(\omega,z) \\
            \hat{a}^{\dagger}_{I}(-\omega,z)
        \end{pmatrix} = \bm{U}^{\text{Dist}}(\omega;z-z_{0})\begin{pmatrix}
            \hat{a}_{S}(\omega,z_0) \\
            \hat{a}^{\dagger}_{I}(-\omega,z_0)
        \end{pmatrix}\nonumber\\&+\int_{z_0}^{z} dz'\bm{U}^{\text{Dist}}(\omega;z-z')\begin{pmatrix}
            \sqrt{\kappa_{S}(\omega)}\hat{b}_{S}(\omega,z') \\
            \sqrt{\kappa_{I}(-\omega)}\hat{b}^{\dagger}_{I}(-\omega,z')
        \end{pmatrix}.
    \end{align}
    From these solution, we can obtain expressions for the second order moments at the end of a lossy nonlinear region of length $L$ starting at the origin w.r.t. vacuum inputs
    \begin{align}
        N^{\text{Dist}}_{S}(\omega)=&N^{\text{Dist,V}}(\omega;L)\nonumber\\&+\kappa_{I}(-\omega)\int_{0}^{L}dz'N^{\text{Dist,V}}(\omega;L-z'),\label{eq:distributed_signal} \\
        N^{\text{Dist}}_{I}(-\omega)=&N^{\text{Dist,V}}(\omega;L)\nonumber\\&+\kappa_{S}(\omega)\int_{0}^{L}dz'N^{\text{Dist,V}}(\omega;L-z'), \\
        M^{\text{Dist}}(\omega) =& M^{\text{Dist,V}}(\omega;L) \nonumber\\&+\kappa_{S}(\omega)\int_{0}^{L}dz'M^{\text{Dist,V}}(\omega;L-z')\label{eq:distributed_cross},
    \end{align}
    where, similarly to the lossless case, we define
    \begin{align}
        N^{\text{Dist,V}}(\omega;z)&\equiv N^{\text{Dist,V}}_{S}(\omega;z) = N^{\text{Dist,V}}_{I}(-\omega;z)\nonumber\\&=\bm{U}^{\text{Dist};S,I}(\omega;z)\left[\bm{U}^{\text{Dist};S,I}(\omega;z)\right]^{*},\\
        M^{\text{Dist, V}}(\omega;z) &= \bm{U}^{\text{Dist};S,S}(\omega;z)\left[\bm{U}^{\text{Dist};I,S}(\omega;z)   \right]^{*},
    \end{align}
    in terms of the matrix elements and explicitly we have that
    \begin{align}
        N^{\text{Dist,V}}(\omega;z)=& e^{-(\kappa_{S}(\omega)+\kappa_{I}(-\omega))z/2}\left|\frac{2\gamma\sin{(\tilde{\nu}(\omega) z/2)}}{\tilde{\nu}(\omega)}\right|^{2} ,\label{eq:Nsvdist} \\
        M^{\text{Dist, V}}(\omega;z)=& 2i\gamma e^{-(\kappa_{S}(\omega)+\kappa_{I}(-\omega))z/2}\left[ \frac{\sin{(\tilde{\nu}(\omega) z/2)}}{\tilde{\nu}(\omega)}  \right]^{*}\nonumber\\ &\cdot\left[ \cos{\left(\frac{\tilde{\nu}(\omega) z}{2}\right)}+i\frac{\tilde{\Sigma}_{K}(\omega)}{\tilde{\nu}(\omega)}\sin{\left(\frac{\tilde{\nu}(\omega) z}{2}\right)}    \right]\label{eq:Mvdist},
    \end{align}
    where we have defined
    \begin{align}
    \tilde{\Sigma}_{K}(\omega) &= \Delta k_{S}(\omega)+\Delta k_{I}(-\omega)+i\frac{\kappa_{S}(\omega)-\kappa_{I}(-\omega)}{2}\\
        \tilde{\nu}(\omega) &= \sqrt{\left[\tilde{\Sigma}_{K}(\omega)\right]^{2}-4|\gamma|^{2}} \label{eq:lossmatching}.
    \end{align}
    Note that if we set $\kappa_{S}=\kappa_{I}=0$ in Eqs.~(\ref{eq:distributed_signal})~--~(\ref{eq:distributed_cross}), we regain the solutions to the lossless case (Eqs.~(\ref{eq:Nv}) and (\ref{eq:Mv_exp})) as one would expect.

    As can be seen from Eqs.~(\ref{eq:distributed_signal}) and (\ref{eq:Nsvdist}), the signal intensity explicitly depends on the idler decay rate via an exponential decay term, the modified $\tilde{\nu}(\omega)$ parameter, and the noise integral contribution. Therefore, idler-only loss affects the signal intensity in the DL model.

    \subsection{Parametrization}\label{sec:param}

    As we are interested in comparing all three models for different levels of idler loss and parametric gain, we need to parametrize these quantities consistently.
    
    By considering the solutions to the DL model's equations of motion with the nonlinear interaction turned off, we find that they are equivalent to a beamsplitter transformation if we define the transmission coefficient as
    \begin{align}\label{eq:kappa_to_trans}
        \eta_{\mu}(\omega) = e^{-\kappa_{\mu}(\omega)L} 
    \end{align}
    for $\mu\in\{S,I\}$. We therefore define the SU(1,1) and IC transmission coefficients via the DL decay rates this way. This parametrization ensures that a change of interaction length in the DL model is mirrored in the SU(1,1) and IC configurations. 
    
    We characterize the level of gain via the peak value of the single-pass signal intensity when no loss is present $N^{P}_{S} = \text{max}_{\omega}\left[N^{\text{V}}_{S}(\omega)\right]$ where ``P'' stands for peak. This peak value occurs when we have perfect phase-matching (i.e. $\Sigma_{K}(\omega)=\tilde{\Sigma}_{K}(\omega)=0$) which gives $N^{P}_{S}=\sinh^{2}{(\gamma L)}$ and we tune the peak value by tuning the interaction strength $\gamma$. Recall that when there is no loss present, all three models give rise to the same single-pass signal intensity (i.e. $N^{\text{V}}_{S}(\omega)$).

	\section{Quantum Fisher Information}
	\label{sec:QFI}
    
	\begin{figure*}
		\includegraphics[width=0.95\linewidth]{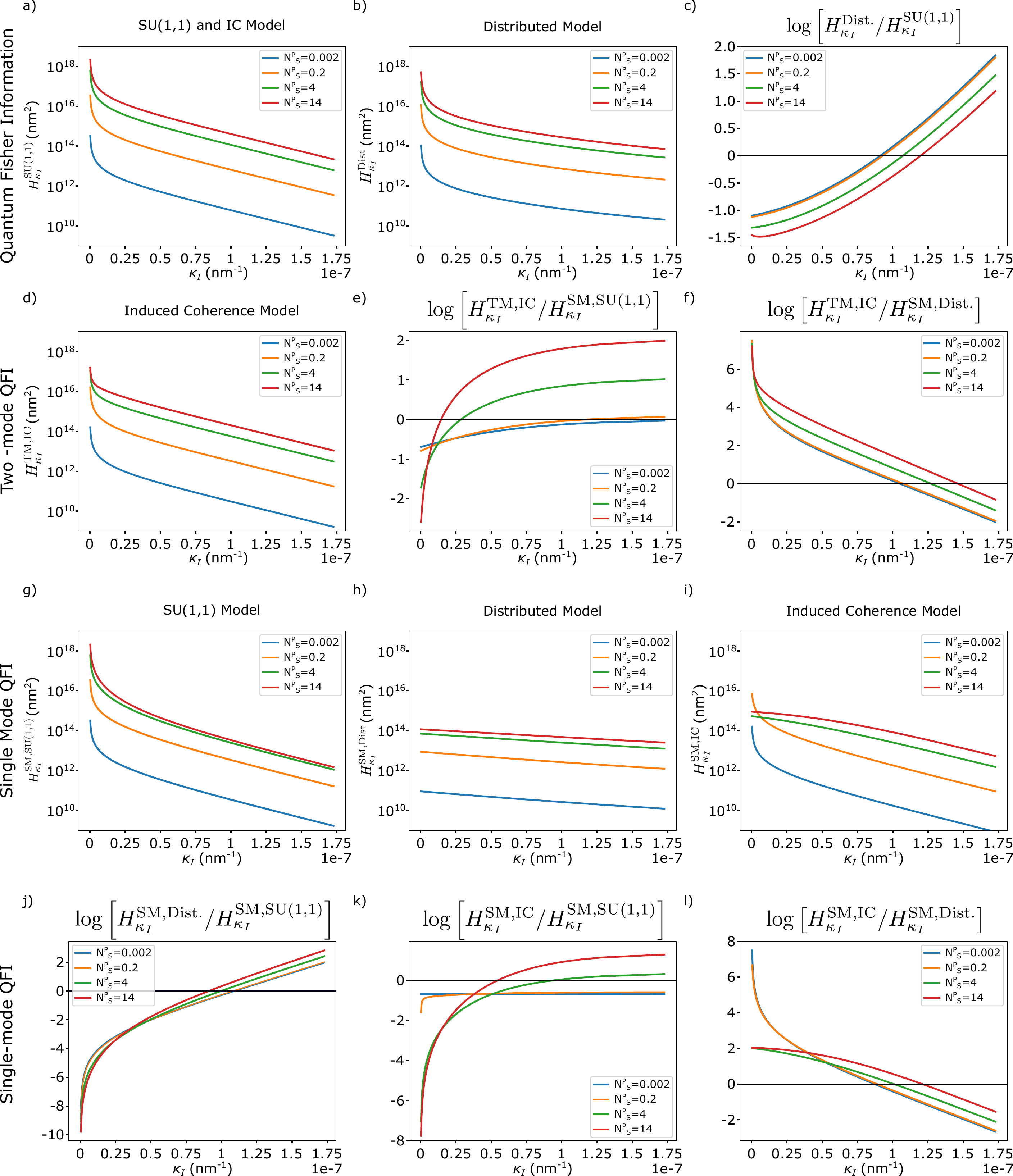}
		\caption{Comparing quantum fisher information for the SU(1,1), DL, and IC models. First row shows the full QFI with access to all modes (c.f. Eq.~(\ref{eq:qfi_twomode})) for the (a) SU(1,1) and IC models, and (b) DL model. (c) shows the logarithm of the ratio of the DL and SU(1,1) QFI. (d) shows the two-mode QFI for the IC model after the idler is traced out (c.f. Eq~(\ref{eq:tm_ic_qfi})). (e) and (f) show the logarithm of the ratio between the two-mode IC QFI and the single-mode QFI of the SU(1,1) and DL respectively. The third row shows the single-mode QFI for the (g) SU(1,1), (h) DL, and (i) IC models. The fourth row shows the logarithm of the ratio of the single-mode QFI between the (j) DL and SU(1,1) models, (k) IC and SU(1,1) models, (l) IC and DL models.  Each figure includes several curves at different levels of gain. We plot the curves as a function of $\kappa_{I}$ and the chosen range corresponds to a transmission coefficient $\eta_{I}\in(99,0.1)\%$. Length of the nonlinear regions is set to $L=40$ mm.}
		\label{fig:qfi_2}
	\end{figure*}

    For a certain parameter, say $\epsilon$, that one wants to estimate, the QFI $H_\epsilon$ is a measure that tells us how much information we can obtain from a certain system (and/or model) concerning said parameter, regardless of the measurement scheme \cite{Helstrom1969,Holevo1982ProbabilisticAS,Braunstein1994QFIseminal,Paris2009QFIseminal}. The QFI defines the quantum Cram\'er-Rao bound \cite{Braunstein1994QFIseminal} which tells us that the error for estimating a given parameter is bounded from below by the inverse of the QFI:
    \begin{align}
        \Delta^{2}\epsilon \geq \frac{1}{H_{\epsilon}}.
    \end{align}
    Therefore, the higher the QFI, the more precise a measurement of the estimation parameter. The QFI is in fact a measure of how close a state $\rho_{\epsilon}$ and $\rho_{\epsilon+d\epsilon}$ are to each other (for a small parameter $d\epsilon$) \cite{Safranek2015multimode}. For general systems, the QFI is defined as \cite{Safranek2015multimode}
    \begin{align}
        H_{\epsilon} = 8 \lim_{d\epsilon\to 0} \frac{1-\sqrt{F(\rho_{\epsilon},\rho_{\epsilon+d\epsilon})}}{d\epsilon^{2}}
    \end{align}
    where $F(\rho_{1},\rho_{2})=\Tr\left(\sqrt{\sqrt{\rho_{1}}\rho_{2} \sqrt{\rho_{1}}}   \right)^2$ is the Uhlmann fidelity \cite{Uhlman1976fidelity}. This general expression can be difficult to solve analytically, however, from it we can obtain a relation between the QFI for two different estimation parameter parametrizations 
    \begin{align}\label{eq:qfi_relation}
        H_{\epsilon} = H_{\epsilon'}\left(\frac{\partial \epsilon'}{\partial \epsilon}   \right)^{2}. 
    \end{align}
    This is exactly the relation one expects for error propagation. Since we can parametrize loss by either the transmission or decay rate (see Eq.~(\ref{eq:kappa_to_trans})), Eq.~(\ref{eq:qfi_relation}) will allow us to easily convert between the QFI for estimating $\eta_{I}$ and $\kappa_{I}$. Reintroducing the frequency dependence, we find that
    \begin{align}
        H_{\kappa_{I}(-\omega)}(\omega) &= H_{\eta_{I}(-\omega)}(\omega)\left(\frac{\partial \eta_{I}(-\omega)}{\partial \kappa_{I}(-\omega)}   \right)^{2} \nonumber\\
        &= H_{\eta_{I}(-\omega)}(\omega) L^{2}\eta^{2}_{I}(-\omega).
    \end{align}
    Note that we can evaluate both QFIs as a function of either the transmission or decay rate.
    
    For Gaussian states, the QFI can be expressed solely in terms of the covariance matrix which in certain cases allows one to find analytic solutions. Depending on the number of modes present, the QFI for Gaussian states simplifies to different expressions \cite{Safranek2015multimode, pinel2013singlemode} 
    
    Although we are motivated by learning about an unmeasured idler-only parameter by measuring the signal, it is still of relevance to study how much information we can gather from each model when we have access to all possible modes. 
    
	\subsection{Access to all modes}
	\label{sec:fullQFI}

    We begin by considering the case where we have access to the idler mode in addition to the signal and ancilla modes. Since the QFI is independent of the measurement scheme, we need to be careful with where we evaluate it in the different configurations. In the SU(1,1) and IC configurations, the second pass through the lossless squeezers does not imprint nor remove any information concerning the idler loss but rather mixes it between the modes: the second pass is thus considered as part of the measurement scheme. One must therefore evaluate the QFI before the second pass at which point the SU(1,1) and IC configurations are equivalent from an information theoretic standpoint. Therefore, we only need to consider the QFI for two-mode Gaussian states.
    
    For two-mode Gaussian states, the QFI can be expressed as~\cite{Safranek2015multimode}
    \begin{widetext}\begin{align}\label{eq:qfi_twomode}
    H_{\epsilon}=& \frac{1}{2(|\Gamma|-1)}\Biggl[ |\Gamma|\text{Tr}\left[\left(\Gamma^{-1}\dot{\Gamma}  \right)^{2} \right] + \sqrt{|1+\Gamma^{2}|}\text{Tr}\left[\left((1+\Gamma^{2})^{-1}\dot{\Gamma}  \right)^{2} \right]   +4(\lambda_{1}^{2}-\lambda_{2}^{2})\left(-\frac{\dot{\lambda}^{2}_{1}}{\lambda_{1}^{4}-1}+\frac{\dot{\lambda}^{2}_{2}}{\lambda_{2}^{4}-1}  \right)\Biggr]
    \end{align}\end{widetext}
    where $|\cdot|$ is the determinant, the derivative w.r.t $\epsilon$ is expressed using dot-notation, $\Gamma=\Omega\sigma$ is the symplectic covariance matrix with $\Omega=\mathbb{1}_{2\times2}\oplus -\mathbb{1}_{2\times2}$ and $\sigma_{i,j} =  \langle \bm{r}_{i}\bm{r}^{\dagger}_{j}+\bm{r}^{\dagger}_{j}\bm{r}_{i}\rangle$ with $\bm{r} =\left( \hat{a}_{S}(\omega),\hat{a}_{I}(-\omega),\hat{a}_{S}^{\dagger}(\omega),\hat{a}_{I}^{\dagger}(-\omega)\right)$. Note that we do not have the $1/2$ factor in the definition of the covariance matrix. $\Gamma$ is a $4\times4$ matrix which has eigenvalues that come in $\pm$ pairs, as such $\lambda_{1,2}$ are the positive eigenvalues of $\Gamma$. Although this formula is quite cumbersome, it is possible to obtain analytic expressions\cite{Adesso2009absformula,Allen2020absformula,Woodworth2020absformula,Sahota2015phaseestimation}.
    
    In the case of idler only loss, it has been shown that the QFI for the SU(1,1) model w.r.t estimation parameter $\eta_{I}(-\omega)$ is\cite{Adesso2009absformula,Allen2020absformula,Nair2018losssensing,Woodworth2020absformula}
    \begin{align}
        H^{\text{SU(1,1)}}_{\eta_{I}(-\omega)}(\omega) = \frac{N^{V}(\omega)}{\eta_{I}(-\omega)\left(1-\eta_{I}(-\omega)  \right)}  = H^{\text{IC}}_{\eta_{I}(-\omega)}(\omega)
    \end{align}
    where $N^{V}(\omega)$ is given by Eq.~(\ref{eq:Nv}). For the rest of the analysis, we consider a specific frequency where the squeezers are phase-matched to maximize their effects. Dropping the frequency argument, the QFI for estimating the decay rate in the SU(1,1) and IC models takes the form
    \begin{align}\label{eq:qfi_su11}
        H^{\text{SU(1,1)/IC}}_{\kappa_{I}} = \frac{L^2N^{P}_{S}}{e^{\kappa_{I}L}-1}.
    \end{align}
    In Fig.~\ref{fig:qfi_2}(a) we plot Eq.~(\ref{eq:qfi_su11}) for different values of gain on a log-scale. The QFI blows up as $\kappa_{I}$ goes to zero and as we increase the decay rate, the QFI decreases linearly (due to log-scaling). 
    
    For the DL model, we numerically evaluate Eq.~(\ref{eq:qfi_twomode}) and show the results in Fig.~\ref{fig:qfi_2}(b) for different levels of gain. Again, the QFI diverges as $\kappa_{I}$ goes to zero and then decreases as we increase the decay rate. However, for the DL model, the QFI does not decrease linearly (on the log-scale) as a function of the decay rate and as we can see, it decays more slowly than for the SU(1,1) model. To attempt to understand this, we obtain an approximate form for the DL QFI which agrees well with the results (see Appendix~\ref{app:derivation} for details on how we obtain the approximate expression)  
    \begin{align}\label{eq:qfi_approx}
        H^{\text{Dist.}}_{\kappa_{I}}\approx\frac{N^{\text{Dist.}}_{S}-N^{\text{Dist.}}_{I}}{\alpha \kappa_{I}^{2}}
    \end{align}
    where $\alpha=1.1$  is a fit parameter. Note that in this expression, there is no exponentially decaying term and so the QFI of the DL model decays much slower than that of the SU(1,1) model.

    In Fig.~\ref{fig:qfi_2}(c) we plot $\log\left[ H^{\text{Dist.}}_{\kappa_{I}}/H^{\text{SU(1,1)}}_{\kappa_{I}}  \right]$ to determine which model has the larger QFI. Initially, we find that $H^{\text{Dist.}}_{\kappa_{I}} < H^{\text{SU(1,1)}}_{\kappa_{I}}$. This is due to the dependence of the QFIs on the intensities since we have that $\left(N^{\text{Dist.}}_{S}-N^{\text{Dist.}}_{I}\right) < N^{V}_{P}$. As we increase the decay rate, we eventually get to a point where $H^{\text{Dist.}}_{\kappa_{I}} > H^{\text{SU(1,1)}}_{\kappa_{I}}$. This is due to the different decay rate scaling. This cross-over depends on the level of gain, however, it occurs in the vicinity of $\kappa_{I}\approx 1\cdot10^{-7}$ $\text{nm}^{-1}$ which corresponds to a transmission rate of $\eta_{I}\approx 1.8\%$. 

    At the level of the QFI when we have access to all modes, we conclude that the SU(1,1) and IC setups are better for absorption estimation unless we have high levels of loss, $\eta_{I}\lesssim 1\%$, at which point the DL setup becomes optimal. Although the QFI tells us which model we can extract the most information from, it does not tell us what measurement scheme will give us said maximal information. In Appendix~\ref{sec:intensity_diff}, we consider the error obtained from an an intensity difference measurement between the signal and the idler for the SU(1,1) and DL models.

	\subsection{No access to idler mode}
	\label{sec:signalQFI}
    
    We now consider the QFI when we do not have access to the idler mode. In this case, the QFI of the SU(1,1) and IC configurations needs to be evaluated after the second pass. For the IC configuration, the ancilla is now relevant to the analysis and after tracing out the idler we are still required to use the formula for two-mode Gaussian states. On the other hand, for the SU(1,1) and DL models we are left with a single-mode thermal Gaussian state after we trace out the idler. For the IC model, we also consider a single-mode operation where we trace out one of the beamsplitter arms after the signal and ancilla modes are mixed. Note that for the IC model, one could also consider the case where the ancilla and signal are not mixed and we only have access to the ancilla (signal). However, in this case, tracing out the signal (ancilla) and idler removes too much information and the QFI is always lower than the single-mode SU(1,1) and DL QFIs.

    \subsubsection{Induced coherence: access to ancilla and signal mode}\label{sec:IC_twomode}
    After we trace out the idler in the IC setup, we are left with a classical two-mode Gaussian state (see Appendix~\ref{sec:cov_mat} for its covariance matrix). The simplified expressions for idler only loss and phase-matched squeezers presented in Sec.~\ref{sec:IC} allow us to obtain an analytic expression for the QFI w.r.t. estimation parameter $\kappa_{I}$. We find that the QFI for the two-mode IC configuration is
    \begin{align}
        H^{\text{TM},\text{IC}}_{\kappa_{I}} &= \frac{L^2 N^{P}_{S} e^{-\kappa_{I}L}(1+N^{P}_{S}(1-e^{-\kappa_{I}L}))}{(1-e^{-\kappa_{I}L})(2+N^{P}_{S}(2-e^{-\kappa_{I}L}))}\nonumber\\
    & = H^{IC}_{\kappa_{I}}\frac{(1+N^{P}_{S}(1-e^{-\kappa_{I}L})}{(2+N^{P}_{S}(2-e^{-\kappa_{I}L}))}\label{eq:tm_ic_qfi}
    \end{align}
    where the superscript ``TM'' refers to the fact that we are considering two-modes and $H^{IC}_{\kappa_{I}}$ is the QFI with access to all modes (Eq.~(\ref{eq:qfi_su11})). This result agrees with those of \cite{volkoff2024radar}. In Fig.~\ref{fig:qfi_2}(d) we show the two-mode QFI w.r.t. $\kappa_{I}$ for the IC model for different levels of gain. From this expression and by comparing the figures, we clearly see that the two-mode IC QFI is less than the full mode QFI. This is to be expected since we are removing information when tracing out the idler. An interesting question to investigate is to see whether or not the addition of the ancilla mode gives an estimation advantage when comparing to the single-mode SU(1,1) and DL models.

    \subsubsection{Single-mode operation}\label{sec:single}

    For single-mode operation, we have that each configuration is in a single-mode thermal Gaussian state (see Appendix~\ref{sec:cov_mat} for covariance matrix). The QFI for single-mode Gaussian states, and for estimation parameter $\epsilon$, takes the form~\cite{pinel2013singlemode}
    \begin{align}
        H^{\text{SM}}_{\epsilon} = \frac{\text{Tr}\left[(\sigma^{-1}\dot{\sigma})^2    \right]}{2(1+P^2)}+ \frac{2\dot{P}^2}{1-P^{4}}
    \end{align}
    where we define the purity $P = |\sigma |^{-1/2}$ and the superscript ``SM'' refers to single-mode. In full generality, we find that the QFI for a single-mode thermal Gaussian state is
    \begin{align}
     H^{\text{SM}}_{\epsilon} = \frac{\dot{N_{S}^2}}{N_{S}(1+N_{S})}.
    \end{align}
    As a function of the signal intensity, error-propagation gives us
    \begin{align}
        \Delta^{2} \epsilon = \frac{\Delta^{2} \hat{N}_{S}}{\dot{N}^{2}_{S}}
    \end{align}
    and as we show in Appendix~\ref{app:var}, $\Delta^{2} \hat{N}_{S} = N_{S}(N_{S}+1)$. We then have that, for our estimation parameter $\epsilon$, 
    \begin{align}
        \Delta^{2} \epsilon = \frac{ N_{S}(N_{S}+1)}{\dot{N}^{2}_{S}} = \frac{1}{H^{\text{SM}}_{\epsilon}}.
    \end{align}
    This tells us that the single-mode QFI is exactly the inverse of error propagation given by a signal intensity measurement for thermal Gaussian states. Therefore, if we only have access to a single mode, an intensity measurement will saturate the quantum Cram\'er-Rao bound. In Figs.~\ref{fig:qfi_2}(g-i) we show the single-mode QFI for the SU(1,1), DL, and IC models respectively. Note that for the single-mode IC configuration, where we choose phases such that $\sin(\frac{\Delta K L}{2}-\Phi) = 1$ in Eq.~(\ref{eq:IC_bbs}), we find that the $N^{\text{IC,BBS}}_{-}$ gives rise to the optimal QFI.

    \subsubsection{Comparison}\label{sec:analysis}

    We can now compare the the QFI for the different configurations when the idler is traced out. In Figs.~\ref{fig:qfi_2}(e) and (f) we consider the logarithm of the ratio of the two-mode IC QFI to the single-mode QFI of the SU(1,1) and DL models respectively. From Fig.~\ref{fig:qfi_2}(e), we see that the IC setup becomes better than the single-mode SU(1,1) but mostly in the high-gain limit. Although the IC setup does become better in the low-gain, the level of loss at which it occurs is at a point where the DL model becomes optimal. When comparing to the DL model, we again see similar behaviour where the DL model becomes better in the high-loss limit in the vicinity of $\kappa_{I}\approx 1\cdot10^{-7}$ $\text{nm}^{-1}$. 

    Comparing the single-mode QFIs (Figs.~\ref{fig:qfi_2}(g-i)) to the two-mode QFIs (Figs.~\ref{fig:qfi_2}(a), (b), and (d)) , we see that the single-mode QFIs can be orders of magnitude lower. This is to be expected since we are tracing out a mode and therefore losing information. In Fig.~\ref{fig:qfi_2}(j) we show the logarithm of the ratio of the single-mode QFI of the SU(1,1) and DL models. We find similar behaviour to the previous cases: the SU(1,1) model has a higher single-mode QFI unless we are in the high-loss regime at which point the single-mode QFI of the DL model is higher. This cross-over occurs again in the vicinity of $\kappa_{I}\approx 1\cdot10^{-7}$ $\text{nm}^{-1}$. In 
    Figs.~\ref{fig:qfi_2}(k) and (l) we show the logarithm of the ratio of the single-mode IC QFI to the single-mode QFI of the SU(1,1) and DL models respectively. The behaviour is very similar to the two-mode IC QFI. The high-gain advantage over the SU(1,1) model is not as good and occurs at higher levels of loss, nearing the point at which the DL model becomes the optimal setup.  
    
    In most of the cases presented above, we find that the SU(1,1) model performs better at estimating absorption until we reach the high-loss regime where $\eta_{I}\lesssim 1\%$ at which point the DL model becomes optimal. In the high-gain regime we do find that there is a window of loss for which the IC setup performs better than the SU(1,1) model when the idler is traced out, however, as we increase loss we also find that the DL model performs better than the IC setup in the vicinity of $\eta_{I}\lesssim 1\%$. 
    
    Experimentally however, one needs to be careful. SU(1,1) and IC systems typically also include insertion loss when entering the second squeezer, which we have not considered here. This will in turn lead to higher level of loss encountered by the modes and skew the results. One thus needs to characterize the insertion loss on a case by case basis to obtain better results for the absorption estimation in an SU(1,1) or IC setup. Although the DL model is only optimal in the high-loss limit, it does not have this additional insertion loss. 
	
	\section{Conclusion}
	\label{sec:conc}

    We have developed and compared three theoretical models for absorption spectroscopy with undetected photons: an SU(1,1) interferometer, an IC configuration, and a DL scheme. Starting from a lossless continuous‑wave model of two‑mode squeezing, we obtained analytic expressions for the second‑order photon‑number moments of the SU(1,1) and IC models. Extending this squeezing model, we derived Heisenberg–Langevin equations for a single source with distributed loss and again obtained analytical expressions for the second‑order photon‑number moments.
    
    To assess the metrological performance of the three configurations, we analyzed their QFI as a function of parametric gain and loss for different detection scenarios—ranging from full multimode access to signal‑only readout. Across most experimentally relevant conditions of moderate gain and losses below 99\%, the SU(1,1) interferometer yields the highest QFI. At high gain and intermediate loss, the IC configuration becomes favourable, while at extreme attenuation (transmission $< 1\%$) the DL model performs best. Practical implementations, however, must also consider additional insertion losses in interferometric setups and the phase‑stabilization requirements of the IC scheme.
    
    By establishing a unified theoretical framework, this work enables quantitative benchmarking of different sensing architectures based on the characteristics of a given PDC source. The approach facilitates the design of integrated quantum sensors optimized for specific absorption regimes, providing clear criteria for selecting the configuration that minimizes estimation uncertainty for a targeted loss range.

    \section*{Acknowledgements}
	MH and NQ acknowledge support from the Minist\`{e}re de l'\'{E}conomie et de l’Innovation du Qu\'{e}bec and the Natural Sciences and Engineering Research Council of Canada. This work has been funded by the European Union's Horizon Europe Research and Innovation Programme under agreement 101070700 project MIRAQLS. FR is member of the Max Planck School of Photonics supported by the German Federal Ministry of Research, Technology and Space (BMFTR), the Max Planck Society, and the Fraunhofer Society. 

    \section*{Data Availability Statement}
    
    Raw data and analysis code are available from the corresponding author upon reasonable request.

	\appendix

	\section{Second order moments derivation}\label{app:moments}
    To derive the expressions of the second order moments given in Sec.~\ref{sec:cw}, we use Eq.~(\ref{eq:matrix_eom}) of the main text and keep everything in terms of the propagator matrix elements. Since this model represents a perfect squeezer where signal and idler are identical(w.r.t. their central frequencies), we only consider the signal intensity and the phase-sensitive correlator. As in the main text, we assume without loss of generality that the nonlinear region spans the region from $z_{0}=0$ to $z=L$. Since the interaction itself is z-independent the equations presented are valid for any segment of equal length(regardless of beginning and endpoint). This is only true for the second order moments, the modes themselves will behave differently depending on the origin(i.e. due to second order dispersion). As in the main text, we label modes before(after) the interaction region with the superscript ``in(out)''. Using Eq.~(\ref{eq:matrix_eom}), we can express the signal intensity, w.r.t. vacuum, as a function of the propagator matrix elements
    \begin{align}\label{eq:A_ns}
        &N^{\text{out}}_{S}(\omega)=\langle\hat{a}^{\dagger}_{S}(\omega,L)\hat{a}_{S}(\omega,L) \rangle_{ \text{vac}}\nonumber\\
        &=\langle\left( \left[\bm{U}^{S,S}(\omega;L)\right]^{*}\hat{a}^{\dagger}_{S}(\omega,0)+ \left[\bm{U}^{S,I}(\omega;L)\right]^{*}\hat{a}_{I}(-\omega,0)\right)
        \nonumber\\&
        \hspace{12pt}\cdot\left( \bm{U}^{S,S}(\omega;L)\hat{a}_{S}(\omega,0)+ \bm{U}^{S,I}(\omega;L)\hat{a}^{\dagger}_{I}(-\omega,0)  \right)\rangle_{ \text{vac}}\nonumber\\
        &= \left[\bm{U}^{S,S}(\omega;L)\right]^{*}\bm{U}^{S,S}(\omega;L)\langle\hat{a}^{\dagger}_{S}(\omega,0)\hat{a}_{S}(\omega,0)  \rangle_{ \text{vac}}\nonumber\\
        &\hspace{12pt}+\left[\bm{U}^{S,S}(\omega;L)\right]^{*}\bm{U}^{S,I}(\omega;L)\langle\hat{a}^{\dagger}_{S}(\omega,0)\hat{a}^{\dagger}_{I}(-\omega,0)  \rangle_{ \text{vac}}\nonumber\\
        &\hspace{12pt}+\left[\bm{U}^{S,I}(\omega;L)\right]^{*}\bm{U}^{S,S}(\omega;L)\langle\hat{a}_{I}(-\omega,0)\hat{a}_{S}(\omega,0)  \rangle_{ \text{vac}}\nonumber\\
        &\hspace{12pt}+\left[\bm{U}^{S,I}(\omega;L)\right]^{*}\bm{U}^{S,I}(\omega;L)\langle\hat{a}_{I}(-\omega,0)\hat{a}^{\dagger}_{I}(-\omega,0)  \rangle_{ \text{vac}}\nonumber\\
        &=\left[\bm{U}^{S,S}(\omega;L)\right]^{*}\bm{U}^{S,S}(\omega;L)N^{\text{in}}_{S}(\omega)\nonumber\\
        &\hspace{12pt}+\left[\bm{U}^{S,S}(\omega;L)\right]^{*}\bm{U}^{S,I}(\omega;L)\left[M^{\text{in}}(\omega)\right]^*\nonumber\\
        &\hspace{12pt}+\left[\bm{U}^{S,I}(\omega;L)\right]^{*}\bm{U}^{S,S}(\omega;L)M^{\text{in}}(\omega)\nonumber\\
        &\hspace{12pt}+\left[\bm{U}^{S,I}(\omega;L)\right]^{*}\bm{U}^{S,I}(\omega;L)\left( 1+ N^{\text{in}}_{I}(-\omega) \right).
    \end{align}
    For vacuum inputs, we have that $N^{\text{in}}_{S}(\omega)=N^{\text{in}}_{I}(\omega)=M^{\text{in}}(\omega)=0$ which gives us the vacuum relation of Eq.~(\ref{eq:Nsv}) in the main text. We can further simplify the full expression above by using these vacuum relations (for the last three terms). Furthermore, by the conservation of the bosonic commutation relation, we find that $\left[\bm{U}^{S,S}(\omega;L)\right]^{*}\bm{U}^{S,S}(\omega;L) = 1+N^{\text{V}}_{S}(\omega)$. Combining everything together gives the final form of Eq.~(\ref{eq:ns_out}) in the main text.

    From a similar expansion, we have that the cross-correlator is
    \begin{align}\label{A_M}
        M^{\text{out}}(\omega)=&\langle\hat{a}_{S}(\omega,L)\hat{a}_{I}(-\omega,L) \rangle_{ \text{vac}}\nonumber\\
        =&\left[\bm{U}^{I,S}(\omega;L)\right]^{*}\bm{U}^{S,S}(\omega;L)\left( 1+N^{\text{in}}_{S}(\omega) \right)\nonumber\\
        &+\left[\bm{U}^{I,I}(\omega;L)\right]^{*}\bm{U}^{S,I}(\omega;L)N_{I}^{\text{in}}(-\omega)\nonumber\\
        &+\left[\bm{U}^{I,S}(\omega;L)\right]^{*}\bm{U}^{S,I}(\omega;L)\left[M^{\text{in}}(\omega)\right]^{*}\nonumber\\
        &+\left[\bm{U}^{I,I}(\omega;L)\right]^{*}\bm{U}^{S,S}(\omega;L)M^{\text{in}}(\omega).
    \end{align}
    Again, for vacuum inputs we obtain the vacuum relation of Eq.~(\ref{eq:Mv}). The vacuum input relations allow us to rewrite the first and third terms. Since the signal and idler operators commute, we can also express Eq.~(\ref{eq:Mv}) as $ M^{\text{V}}(\omega)=\left[\bm{U}^{I,I}(\omega;L)\right]^{*}\bm{U}^{S,I}(\omega;L)$, allowing us to re-express the second term. Finally, from the analytic expressions for the matrix elements, we find that $\left[\bm{U}^{I,I}(\omega;L)\right]^{*}\bm{U}^{S,S}(\omega;L)=-\frac{\gamma^{*}}{\gamma}\left[M^{\text{V}}(\omega)\right]^{2}/N^{\text{V}}(\omega)$.

	\section{Approximate solution to DL QFI}\label{app:derivation}
\begin{figure}[ht]
		\includegraphics[width=1\linewidth]{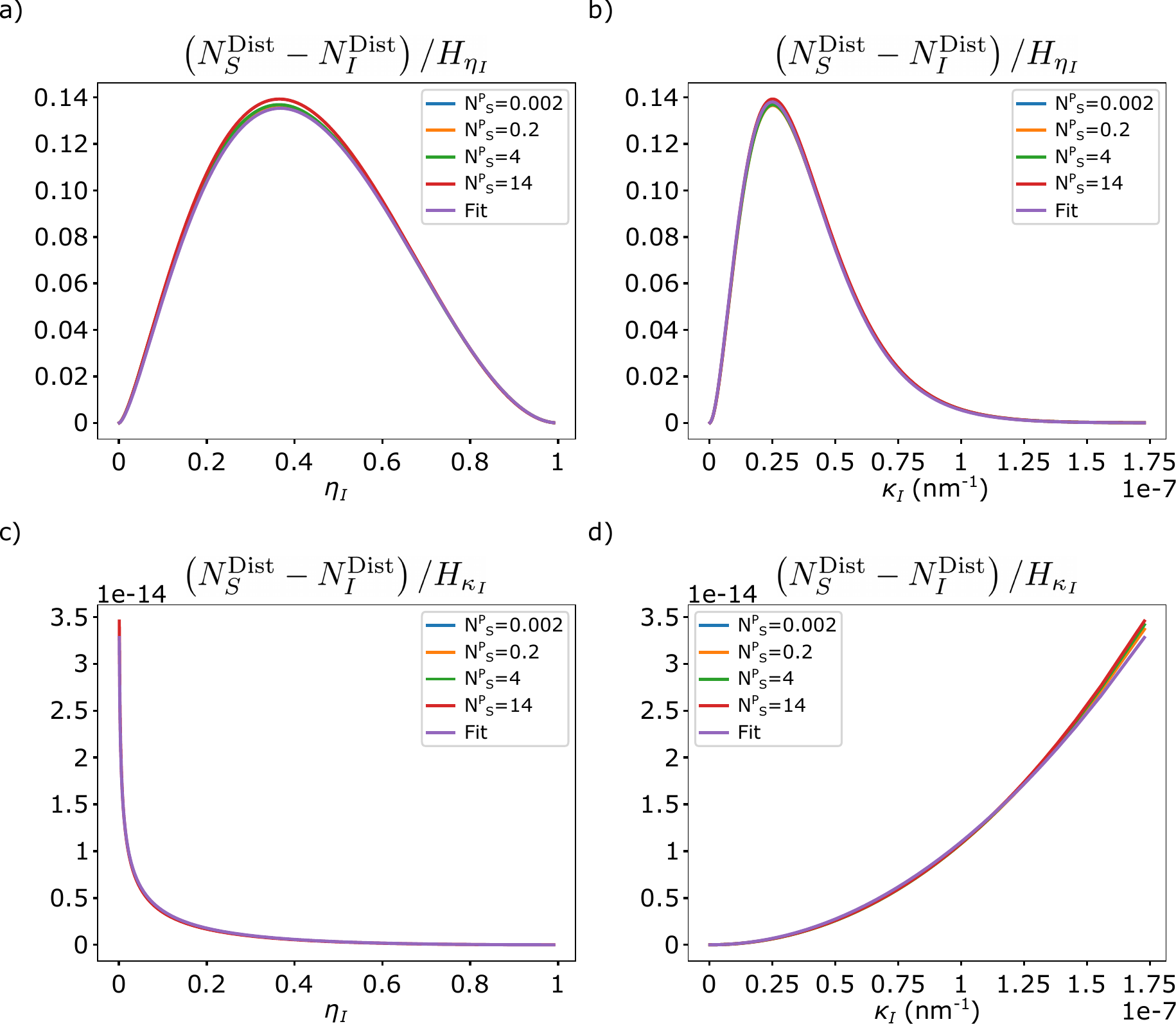}
		\caption{Inverse ratio $\left( N^{\text{Dist.}}_{S}(\omega)-N^{\text{Dist.}}_{I}(-\omega)  \right)/H_{\epsilon}$ for different parametrizations for the DL model. The QFI, $H_{\epsilon}$, is evaluated numerically at a frequency where we are phase-matched. For (a) and (b), the estimation parameter is taken to be $\eta_{I}$ and we plot the inverse ratio as a function $\eta_{I}$ and $\kappa_{I}$ respectively. For (c) and (d), the estimation parameter is taken to be $\kappa_{I}$ and we plot the inverse ratio as a function $\eta_{I}$ and $\kappa_{I}$ respectively. For each plot we include curves at different levels of gain which is determined by $N^{P}_{S}$. Each plot also includes the approximate function, labelled as ``Fit'', with the fit parameter $\alpha=1.1$. Using a least-squared fit, we find an average value of $R^{2}=0.998$ when comparing the numerical solutions to our approximate form. Chosen range corresponds to a transmission coefficient $\eta_{I}\in(99,0.1)\%$. Length of the nonlinear regions is set to $L=40$ mm.}
		\label{fig:approx_form}
	\end{figure}

    In this appendix, we discuss how we came to the approximate form for the DL model QFI. We were initially motivated by the fact that for the SU(1,1) system, if we divide the QFI by the signal intensity we are left with a function that depends only on the estimation parameter (i.e. is independent of frequency, gain, or any other model-dependent parameter). As such, we considered different ratios between the DL QFI and different intensity or derivatives of intensity w.r.t the estimation parameter combinations. We found that dividing by the intensity difference $N^{\text{Dist.}}_{S}(\omega)-N^{\text{Dist.}}_{I}(-\omega)$ gave a fairly gain-independent result. In Fig.~\ref{fig:approx_form} we show the inverse ratio (i.e.$\left( N^{\text{Dist.}}_{S}(\omega)-N^{\text{Dist.}}_{I}(-\omega)  \right)/H_{\epsilon}$) for four different parametrization. We choose the decay-rate $\kappa_{I}$ such that $\eta_{I}=e^{-\kappa_{I}L}\in(0.99,0.001)$.
    
    In Figs.~\ref{fig:approx_form}(a) and (b), we plot the inverse ratio when the estimation parameter is $\eta_{I}$ as a function of $\eta_{I}$ and $\kappa_{I}$ respectively. Similarly, in Figs.~\ref{fig:approx_form}(c) and (d), we plot the inverse ratio when the estimation parameter is $\kappa_{I}$ as a function of $\eta_{I}$ and $\kappa_{I}$ respectively. For each plot, we consider different levels of gain and we also include the approximate form. 
    
    We determined the approximate form by looking at the shape of the curves in Figs.~\ref{fig:approx_form}(c) and (d). More specifically, from Fig.~\ref{fig:approx_form}(d) we posited that
    \begin{align}
        \frac{ N^{\text{Dist.}}_{S}- N^{\text{Dist.}}_{I}}{H^{Dist}_{\kappa_I}(\kappa_{I})} \approx  \alpha \cdot \kappa^{2}_{I}
    \end{align}
    where $\alpha$ is a fit parameter. For the curves shown in Fig.~\ref{fig:approx_form}, a value of $\alpha=1.1$ was chosen and gave good agreement for all parametrizations. Indeed, when considering a least-squares fitting, we obtain an average $R^{2}=0.998$ for our approximation. 

	\section{Variance calculations}\label{app:var}

    In this appendix we show the explicit calculations for the variance of both the signal intensity and the intensity difference. We omit frequency dependence, but it is understood that all modes are evaluated at the same frequency. The variance for the signal intensity(and similarly for the idler intensity), w.r.t. vacuum inputs, is given by
    \begin{align}
        \Delta^{2}\hat{N}_{S} &= \langle \hat{N}_{S}^2 \rangle_{\text{vac}}-\langle \hat{N}_{S} \rangle_{\text{vac}}^2 \nonumber\\
        &=\langle \hat{a}^{\dagger}_{S}\hat{a}_{S}\hat{a}^{\dagger}_{S}\hat{a}_{S} \rangle_{\text{vac}}-\langle \hat{a}^{\dagger}_{S}\hat{a}_{S} \rangle_{\text{vac}}^2\nonumber\\
        &= \langle \hat{a}^{\dagger}_{S}\hat{a}^{\dagger}_{S}\hat{a}_{S}\hat{a}_{S} \rangle_{\text{vac}}-\langle \hat{a}^{\dagger}_{S}\hat{a}_{S} \rangle_{\text{vac}}-\langle \hat{a}^{\dagger}_{S}\hat{a}_{S} \rangle_{\text{vac}}^2\nonumber\\
        &=2\left(\langle \hat{a}^{\dagger}_{S}\hat{a}_{S} \rangle_{\text{vac}}\right)^2-\langle \hat{a}^{\dagger}_{S}\hat{a}_{S} \rangle_{\text{vac}} -\langle \hat{a}^{\dagger}_{S}\hat{a}_{S} \rangle_{\text{vac}}^2\nonumber\\
        &=N_{S}\left( N_{S} +1\right) 
    \end{align}
     where in the third line we used the commutation relation for bosonic modes, in the fourth line we used Wick's Theorem for two-mode Gaussian states where $\langle \hat{a}_{S}\hat{a}_{S} \rangle_{\text{vac}}=0$, and finally we identify $N_{S}=\langle \hat{N}_{S} \rangle_{\text{vac}}$. Similarly for the next derivation, we identify $N_{I}=\langle \hat{N}_{I} \rangle_{\text{vac}}$ and $M=\langle a_{S}a_{I}\rangle_{\text{vac}}$.

	\begin{figure*}[ht!]
		\includegraphics[width=1\linewidth]{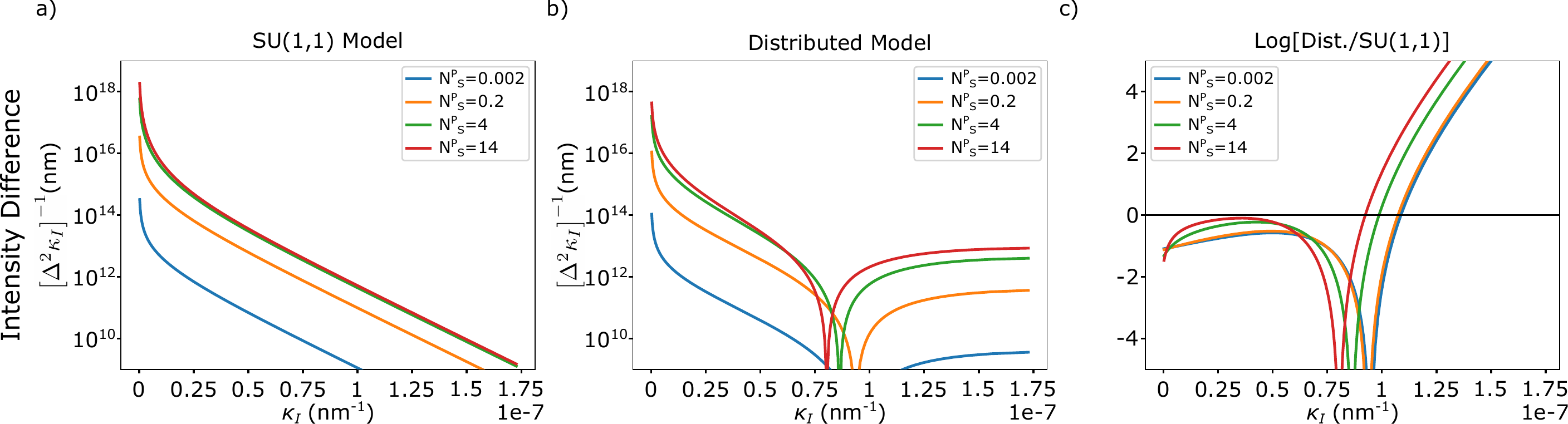}
		\caption{Inverse error obtained from an intensity difference measurement (c.f. Eq.~(\ref{eq:inverse_error})) for the (a) SU(1,1) and (b) DL model. (c) shows the logarithm of the ratio of the inverse errors. Each figure includes several curves at different levels of gain. We plot the curves as a function of $\kappa_{I}$ and the chosen range corresponds to a transmission coefficient $\eta_{I}\in(99,0.1)\%$. Length of the nonlinear regions is set to $L=40$ mm.}
		\label{fig:intensity_diff}
	\end{figure*}

     Having expressed the variance for a single mode, we now consider the variance for the intensity difference
     \begin{align}
         \Delta^{2}\left(\hat{N}_{S}-\hat{N}_{I} \right)=&\langle \left(\hat{N}_{S}-\hat{N}_{I} \right)^2\rangle_{\text{vac}}-\langle \left(\hat{N}_{S}-\hat{N}_{I} \right)\rangle^{2}_{\text{vac}}\nonumber\\
         =& \langle \hat{N}_{S}^2 \rangle_{\text{vac}}-\langle \hat{N}_{S} \rangle_{\text{vac}}^2+\langle \hat{N}_{I}^2 \rangle_{\text{vac}}-\langle \hat{N}_{I} \rangle_{\text{vac}}^2\nonumber\\
         &-2\langle \hat{N}_{S}\hat{N}_{I} \rangle_{\text{vac}}-2N_{S}N_{I}\nonumber\\
         =& \Delta^{2}\hat{N}_{S} +\Delta^{2}\hat{N}_{I}-2\left(\langle \hat{N}_{S}\hat{N}_{I} \rangle_{\text{vac}}-N_{S}N_{I}  \right).
     \end{align}
      Invoking Wick's theorem for two-mode Gaussian states again, we have that
      \begin{align}
          \langle \hat{N}_{S}\hat{N}_{I} \rangle_{\text{vac}} &=\langle \hat{a}^{\dagger}_{S}\hat{a}_{S}\hat{a}^{\dagger}_{I}\hat{a}_{I}\rangle_{\text{vac}}=\langle \hat{a}^{\dagger}_{S}\hat{a}^{\dagger}_{I}\hat{a}_{S}\hat{a}_{I}\rangle_{\text{vac}}\nonumber\\
          &=\langle \hat{a}^{\dagger}_{S}\hat{a}^{\dagger}_{I} \rangle_{\text{vac}}\langle \hat{a}_{S}\hat{a}_{I} \rangle_{\text{vac}} +\langle \hat{a}^{\dagger}_{S}\hat{a}_{S}\rangle_{\text{vac}} \langle\hat{a}^{\dagger}_{I}\hat{a}_{I}\rangle_{\text{vac}}\nonumber\\
          &= |M|^{2} + N_{S}N_{I}
      \end{align}
      where we have used the fact that $\langle \hat{a}^{\dagger}_{S}\hat{a}_{I} \rangle_{\text{vac}}=0$. We then have that the variance for the intensity difference is
      \begin{align}
          \Delta^{2}\left(\hat{N}_{S}-\hat{N}_{I} \right) =\Delta^{2}\hat{N}_{S} +\Delta^{2}\hat{N}_{I}-2|M|^{2}.
      \end{align}
      We can use these formulas to express the different variances after all modular evolutions mentioned in the main text (e.g. nonlinear element, beamsplitter element).

    
    \section{Intensity difference measurement}\label{sec:intensity_diff}


    As a function of the intensity difference, the error for estimating the decay rate is
    \begin{align}\label{eq:inverse_error}
        \Delta^{2}\kappa_{I}= \frac{\Delta^{2}(\hat{N}_{S}-\hat{N}_{I})}{ \left|\frac{\partial (N_{S}-N_{I})}{\partial \kappa_{I}}\right|^2}
    \end{align}
    where
    \begin{align}
        \Delta^{2}(\hat{N}_{S}-\hat{N}_{I}) = N_{S}\left(N_{S}+1 \right)+N_{I}\left(N_{I}+1 \right) -2|M|^{2}
    \end{align}
    in terms of the second order moments (see Appendix~\ref{app:var} for detailed calculations). Since we want to compare the error obtained from an intensity difference measurement to the QFI, in Figs.~\ref{fig:intensity_diff}(a) and (b) we plot the inverse error, $\left[ \Delta^{2}\kappa_{I} \right]^{-1}$, for the SU(1,1) and DL models respectively.

    For both models, we see that this measurement scheme does not saturate the Cram\'er-Rao bound, however, the behaviours are very similar to the QFI with access to all modes. One can make an intensity difference measurement saturate the QFI by passing the signal through an optimized electric gain~\cite{Woodworth2020absformula}. For the SU(1,1) model, Fig.~\ref{fig:intensity_diff}(a), the inverse error diverges as the decay rate goes to zero and decreases linearly (on the log-scale) as a function of the decay rate. The slope of this linear decrease is higher than that of the QFI. For the DL model, Fig.~\ref{fig:intensity_diff}(b), the behaviour has some differences. We find that the derivative of the intensity difference vanishes at some point which causes sharp dips in the inverse error curves. Away from that point, however, we see that as we increase the decay rate, the inverse error does not decay as quickly as in the SU(1,1) model. In Fig.~\ref{fig:intensity_diff}(c) we plot the logarithm of the ratios of the inverse error and again we find similar behaviour to that of the QFI. The SU(1,1) model performs better until the decay rate is in the vicinity of $\kappa_{I}\approx 1\cdot10^{-7}$ $\text{nm}^{-1}$, which corresponds to a transmission rate of $\eta_{I}\approx 1.8\%$, at which point the DL model performs better.

    We omit the intensity difference measurement for the IC configuration as there are several different ways one could do such a measurement. However, unlike the SU(1,1) and DL configurations, an intensity difference measurement in the IC setup will not make use of all the possible information: one of the signals and/or beamsplitter arms is ignored. Therefore, it is not possible to saturate the QFI since some information is lost and/or ignored.

    
    \section{Covariance matrices}\label{sec:cov_mat}


    In this appendix, we include the different covariance matrices used for the different QFI calculations. 

    When we have access to all modes, the covariance matrix for all three configurations (SU(1,1), IC, and DL) is that of a two-mode squeezed state and takes the form
    \begin{align}
        \sigma^{\text{All Modes}} = \begin{psmallmatrix}
            2N_{S}+1& 0 & 0 & 2M \\
            0 & 2N_{I}+1 & 2M & 0 \\
            0 & 2M^{*} & 2N_{S}+1 & 0\\
            2M^{*}& 0 & 0 & 2N_{I}+1 
    \end{psmallmatrix}.
    \end{align}
    For the IC configuration, when we trace out the idler, we find that the two-mode covariance matrix is
    \begin{align}
        \sigma^{\text{IC}} = \begin{psmallmatrix}
        2N^{\text{IC}}_{S}+1 & 2\left[N^{\text{IC}}_{S,A}\right]^{\dagger} & 0 & 0  \\
        2N^{\text{IC}}_{S,A} & 2N^{\text{IC}}_{A}+1  & 0 & 0 \\
        0 & 0 & 2N^{\text{IC}}_{S}+1 & 2N^{\text{IC}}_{S,A}  \\
        0 & 0  & 2\left[N^{\text{IC}}_{S,A} \right]^{\dagger} & 2N^{\text{IC}}_{A}+1
        \end{psmallmatrix}.
    \end{align}
    When we are left with only a single mode for all configurations (tracing out the idler in the SU(1,1) and DL models, and additionally tracing out one of the beamsplitter arms in the IC model) we are left with single-mode thermal Gaussian states. The covariance matrices for these cases take the form
    \begin{align}
        \sigma = \begin{pmatrix}
            2N_{S}+1 & 0  \\
            0  & 2N_{S}+1
        \end{pmatrix}.
    \end{align}
    From these covariance matrices, we are able to obtain the expressions for the different QFIs presented in the main text.

\bibliography{main}

\end{document}